\documentclass[superscriptaddress,aps,prd,nofootinbib,reprint]{revtex4-2}

\usepackage{graphicx}
\usepackage[colorlinks,allcolors=blue]{hyperref} 
\usepackage{amsfonts,amsmath,amssymb,tensor}
\usepackage{mathtools}
\usepackage{braket}
\usepackage{bm}
\usepackage{enumerate}

\begin{document}

\title{Isotropic cosmic birefringence from early dark energy}

\author{Kai Murai}
\affiliation{ICRR, The University of Tokyo, Kashiwa, 277-8582, Japan}
\affiliation{Kavli IPMU (WPI), UTIAS, The University of Tokyo, Kashiwa, 277-8583, Japan}
\author{Fumihiro Naokawa}
\affiliation{Department of Physics, Graduate School of Science, The University of Tokyo, Bunkyo-ku, Tokyo 113-0033, Japan}
\affiliation{Research Center for the Early Universe, The University of Tokyo, Bunkyo-ku, Tokyo 113-0033, Japan}
\author{Toshiya Namikawa}
\affiliation{Kavli IPMU (WPI), UTIAS, The University of Tokyo, Kashiwa, 277-8583, Japan}
\author{Eiichiro Komatsu}
\affiliation{Max Planck Institute for Astrophysics, Karl-Schwarzschild-Str. 1, 85748 Garching, Germany}
\affiliation{Kavli IPMU (WPI), UTIAS, The University of Tokyo, Kashiwa, 277-8583, Japan}

\begin{abstract}
    A tantalizing hint of isotropic cosmic birefringence has been found in the $E B$ cross-power spectrum of the cosmic microwave background (CMB) polarization data with a statistical significance of $3\sigma$. A pseudoscalar field coupled to the CMB photons via the Chern-Simons term can explain this observation. The same field may also be responsible for early dark energy (EDE), which alleviates the so-called Hubble tension. Since the EDE field evolves significantly during the recombination epoch, the conventional formula that relates $E B$ to the difference between the $E$- and $B$-mode auto-power spectra is no longer valid. Solving the Boltzmann equation for polarized photons and the dynamics of the EDE field consistently, we find that currently favored parameter space of the EDE model yields a variety of shapes of the $EB$ spectrum, which can be tested by CMB experiments. 
\end{abstract}

\maketitle

\section{Introduction}
\label{sec: intro}

Observations and interpretation of the cosmic microwave background (CMB) temperature and polarization anisotropies have established the standard cosmological model, the $\Lambda$ Cold Dark Matter ($\Lambda$CDM) model~\cite{Komatsu:2014ioa,Planck:2018vyg,Polarbear:2020lii,ACT:2020gnv,SPT-3G:2021eoc,BICEP:2021xfz,SPIDER:2021ncy}.
There are, however, several observational hints for new physics beyond the $\Lambda$CDM model and the standard model of elementary particles and fields~\cite{Abdalla:2022yfr}. In this paper we connect two such hints: the so-called ``cosmic birefringence'' \cite{Komatsu:2022nvu} and ``Hubble tension'' \cite{Bernal:2016gxb}.

Cosmic birefringence is a rotation of the plane of linear polarization of photons~\cite{Carroll:1989vb,Carroll:1991zs,Harari:1992ea}, which induces the cross-correlation between parity-even $E$ and parity-odd $B$ modes in the CMB polarization field~\cite{Lue:1998mq}.
A tantalizing hint of cosmic birefringence has been found by the recent analyses of the {\sl WMAP} and {\sl Planck} data~\cite{Minami:2020odp,Diego-Palazuelos:2022dsq,Eskilt:2022wav,Eskilt:2022cff}. The reported signal is isotropic and consistent with a rotation angle of $\beta= 0.342^{\circ\, + 0.094^\circ}_{\phantom{\circ\,} - 0.091^\circ}$ (68\% C.L.), independent of the photon frequency~\cite{Eskilt:2022cff}.

A pseudoscalar ``axionlike'' field can explain the observed isotropic cosmic birefringence signal~\cite{Carroll:1998zi,Finelli:2008jv,Arvanitaki:2009fg,Panda:2010uq,Fedderke:2019ajk,Fujita:2020aqt,Fujita:2020ecn,Takahashi:2020tqv,Mehta:2021pwf,Nakagawa:2021nme,Choi:2021aze,Obata:2021nql,Yin:2021kmx,Gasparotto:2022uqo,Lin:2022niw}. In general, a pseudoscalar field can couple to photons via the Chern-Simons term. The known example in the standard model of elementary particles and fields is a neutral pion~\cite{Weinberg:1996kr}.
The temporal and spatial difference in the pseudoscalar field values induces cosmic birefringence independent of the photon frequency.
The evolution of a homogeneous pseudoscalar field induces the isotropic signal.

The Hubble tension refers to the discrepancy between the local and early-universe measurements of the Hubble constant, $H_0$~\cite{Bernal:2016gxb,Riess:2021jrx}. Among the proposed models to alleviate this tension (see Refs.~\cite{DiValentino:2021izs,Schoneberg:2021qvd} for reviews) are early dark energy (EDE) models~\cite{Karwal:2016vyq,Poulin:2018dzj,Poulin:2018cxd,Agrawal:2019lmo,Smith:2019ihp,Ye:2020btb,Rezazadeh:2022lsf}, which utilize an additional energy component to modify the value of $H_0$ inferred from CMB observations.
The EDE field behaves like dark energy in the early universe and starts to oscillate in the pre-recombination epoch. In this paper we focus on an EDE model with a potential of $V(\phi) \propto [1 - \cos (\phi/f)]^n$ with $n > 1$~\cite{Caldwell:2017chz,Poulin:2018cxd}.
Such a potential may be generated for an axionlike field by higher-order instanton corrections~\cite{Abe:2014xja,Kappl:2015esy,Choi:2015aem} or a combination of separate non-perturbative effects~\cite{McDonough:2022pku}.

Several authors estimated the EDE parameters such as the maximum energy-density fraction of EDE, $f_\mathrm{EDE}$, from the current cosmological datasets
and discussed the viability of EDE~\cite{Smith:2019ihp,Hill:2020osr,Ivanov:2020ril,DAmico:2020ods,Murgia:2020ryi,Smith:2020rxx,Seto:2021xua,Hill:2021yec,Poulin:2021bjr,Simon:2022adh}.
The authors of Ref.~\cite{Herold:2021ksg} analyzed the EDE model with $n = 3$ using a profile likelihood to avoid the volume effect on marginalization with the Markov Chain Monte Carlo (MCMC) method, and obtained $f_\mathrm{EDE} = 0.072\pm 0.036$ at 68\% C.L.\footnote{In MCMC, the posterior distribution of parameters is derived from the number of Monte Carlo steps spent in the parameter space. When some parameters are unconstrained, the MCMC ``wastes'' many steps and enhances the posterior probability in that parameter space. In the case of EDE, this occurs in a dramatic way: when $f_\mathrm{EDE}$ goes to zero, all other EDE parameters are unconstrained. This enhances the parameter volume at $f_\mathrm{EDE}=0$, yielding only an upper bound on $f_\mathrm{EDE}$ as found by the previous studies \cite{Hill:2020osr,Ivanov:2020ril,DAmico:2020ods}.}

In this paper we consider the case where the EDE field has a Chern-Simons coupling to photons. Whereas the previous study~\cite{Capparelli:2019rtn} was focused on fluctuations in cosmic birefringence from EDE, we focus on \textit{isotropic} cosmic birefringence. To this end we 
extend the linear Boltzmann solver developed in Ref.~\cite{Nakatsuka:2022epj} and evaluate the $E B$ power spectrum for fixed values of $f_\mathrm{EDE}$ with the best-fitting parameters of the EDE model given in Ref.~\cite{Herold:2021ksg}.

The rest of this paper is organized as follows.
We review the physics of cosmic birefringence induced by an axionlike field in Sec.~\ref{sec: cosmic birefringence}. We explain the EDE model in Sec.~\ref{sec: EDE}. We present the main results of this paper in Sec.~\ref{sec: CB from EDE}. We extend the implementation of cosmic birefringence in the linear Boltzmann solver by including all the relevant effects such as the energy density of $\phi$ in the Friedmann equation and the gravitational lensing effect, which were ignored in the previous work~\cite{Nakatsuka:2022epj}. Thus, this is the first work to fully and self-consistently calculate cosmic birefringence from an axionlike field. We show the resulting $E B$ power spectra, which exhibit complex shapes unique to the EDE model and can be distinguished from the simplest form of cosmic birefringence with a constant rotation angle by future CMB experiments. We summarize our findings and conclude in Sec.~\ref{sec: summary}.

\section{Isotropic cosmic birefringence}
\label{sec: cosmic birefringence}

We consider an axionlike field, $\phi$, coupled to photons through the Chern-Simons term.
The Lagrangian density is written as
\begin{align}
    \mathcal{L}
    =
    - \frac{1}{2} \left( \partial_\mu \phi \right)^2
    - V(\phi)
    - \frac{1}{4} F_{\mu\nu} F^{\mu\nu}
    - \frac{1}{4} g \phi F_{\mu\nu} \tilde{F}^{\mu\nu}
    \, ,
\end{align}
where $V(\phi)$ is the potential which is specified in Sec.~\ref{sec: EDE}, $F_{\mu \nu}$ is the field strength tensor of the photon field, $\tilde{F}^{\mu \nu}$ is its dual, and $g$ is the Chern-Simons coupling constant of mass dimension $-1$.

When $\phi$ varies slowly compared to the photon frequency,
the dispersion relations of photons are modified as~\cite{Carroll:1989vb,Carroll:1991zs,Harari:1992ea}
\begin{align}
    \omega_\pm
    \simeq
    k \mp \frac{g}{2} \left(
        \frac{\partial \phi}{\partial t} + \frac{\bm{k}}{k} \cdot \bm{\nabla} \phi
    \right)
    =
    k \mp \frac{g}{2} \frac{\mathrm{d} \phi}{\mathrm{d} t}
    \, ,
    \label{eq:disp}
\end{align}
where $\mathrm{d}/\mathrm{d}t$ denotes a total derivative along the photon trajectory, and $+$ and $-$ correspond to the right- and left-handed circular polarization states of photons, respectively, in right-handed coordinates with the $z$ axis taken in the direction of propagation of photons.

In this paper we focus on isotropic cosmic birefringence and ignore its anisotropy, which is not found yet~\cite{Contreras:2017sgi,Namikawa:2020ffr,SPT:2020cxx,Gruppuso:2020kfy,Bortolami:2022whx}. See Refs.~\cite{Li:2008tma,Pospelov:2008gg,Caldwell:2011pu,Zhao:2014yna,Agrawal:2019lkr,Zhai:2020vob,Jain:2021shf,Greco:2022ufo,Kitajima:2022jzz,Jain:2022jrp} for study on anisotropic cosmic birefringence.

The helicity-dependent dispersion relation given in Eq.~\eqref{eq:disp} induces a rotation of the plane of linear polarization. The rotation angle from a given time $t$ to the present time $t_0$ is given by~\cite{Carroll:1989vb,Carroll:1991zs,Harari:1992ea}
\begin{align}
    \beta(t)
    =
    - \frac{1}{2} \int_t^{t_0} \mathrm{d}\tilde{t} \, (\omega_+ - \omega_-) 
    =
    \frac{g}{2} \left[ 
        \phi(t_0) - \phi(t)
    \right]
    \, .
\end{align}
Here, we use the CMB convention for the position angle of linear polarization, \textit{i.e.}, $\beta > 0$ represents a clockwise rotation of linear polarization in the sky.
The rotation angle depends only on the difference in the axionlike field values between the emission and observation and does not depend on the evolution history~\cite{Harari:1992ea}.

Next, we discuss the effect of cosmic birefringence on the CMB polarization field (see Ref.~\cite{Komatsu:2022nvu} for a review).
In this paper we focus on scalar-mode perturbations and ignore tensor modes.
The evolution of the CMB polarization field is described by the Boltzmann equation~\cite{Kosowsky:1994cy}.
We define the Fourier transform of the Stokes parameters of linear polarization, $Q\pm iU$, by $_{\pm 2}\Delta_P(\eta, q, \mu)$.
Here, $\eta$ is the conformal time,
$\bm{q}$ is a wave vector in the Fourier space, and $\mu \equiv \bm{q} \cdot \bm{k}/(q k)$ parameterizes the angle between $\bm{q}$ and the photon momentum $\bm{k}$.
The Boltzmann equation for $_{\pm 2}\Delta_P(\eta, q, \mu)$ is given by~\cite{Finelli:2008jv,Liu:2006uh,Gubitosi:2014cua}
\begin{align}
    _{\pm 2} \Delta_P' + i q \mu \, _{\pm 2} \Delta_P
    =
    &\tau' \left[ 
        - _{\pm 2} \Delta_P 
        + \sqrt{\frac{6 \pi}{5}} \, _{\pm2} Y_2^0 (\mu) \Pi(\eta, q)
    \right]
    \nonumber\\
    &\pm 2 i \beta' \, _{\pm 2} \Delta_P
    \, ,
\end{align}
where $_{\pm 2} Y_l^m$ is the spin-2 spherical harmonics, $\Pi(\eta, q)$ is the polarization source term~\cite{Zaldarriaga:1996xe}, and $\tau' \equiv a(\eta) n_e(\eta) \sigma_T$ is the differential optical depth with the Thomson scattering cross section $\sigma_T$ and the number density of electrons $n_e$.
Here, the prime denotes the derivative with respect to $\eta$.
It is convenient to expand $_{\pm 2}\Delta_P$ with the spin-2 spherical harmonics as
\begin{align}
    _{\pm 2} \Delta_P (\eta, q, \mu)
    \equiv 
    \sum_l i^{-l} \sqrt{4 \pi (2l + 1)} 
    \,_{\pm 2} \Delta_{P, l}(\eta, q) \,_{\pm 2} Y_l^0(\mu)
    \, .
\end{align}
Then, we obtain the formal solution to the Boltzmann equation as
\begin{align}
    _{\pm 2} \Delta_{P, l} (\eta_0, q)
    =
    -\frac{3}{4} \sqrt{\frac{(l + 2)!}{(l - 2)!}}
    \int_0^{\eta_0} \mathrm{d} \eta \, 
    & \tau' e^{-\tau(\eta)} \Pi(\eta, q)
    \nonumber\\
    \times & \frac{j_l(x)}{x^2} e^{\pm 2 i \beta(\eta)}
    \, ,
    \label{eq: Solution for Boltzmann}
\end{align}
where $j_l(x)$ is the spherical Bessel function with $x \equiv q (\eta_0 - \eta)$ and $\tau(\eta) \equiv \int^{\eta_0}_\eta \mathrm{d} \eta_1 \, \tau' (\eta_1)$.

To discuss the parity violation, we work with parity eigenstates of the CMB polarization, $E$ and $B$ modes~\cite{Zaldarriaga:1996xe,Kamionkowski:1996ks}.
In this basis, parity violation is imprinted in the $E B$ correlation~\cite{Lue:1998mq}.
We write the coefficients of $E$ and $B$ modes as
\begin{align}
    \Delta_{E, l}(q) \pm i \Delta_{B, l}(q)
    \equiv
    - \,_{\pm 2}\Delta_{P, l}(\eta_0, q)
    \, .
\end{align}
Cosmic birefringence, $\beta \neq 0$, induces the imaginary part of $_{\pm 2}\Delta_{P, l}$, or equivalently $B$ modes.
Using these coefficients, we obtain the polarization power spectra as
\begin{align}
    C_l^{X Y}
    =
    4 \pi \int \mathrm{d} (\ln q) \,
    \mathcal{P}_s(q) \Delta_{X, l}(q) \Delta_{Y, l}(q)
    \, ,
    \label{eq:cl}
\end{align}
where $\mathcal{P}_s (q)$ is the primordial scalar curvature power spectrum, and $X, Y = E$ or $B$.

If $\beta$ is independent of $\eta$, the coefficients of $E$ and $B$ modes are given by
\begin{align}
    \Delta_{E, l} \pm i \Delta_{B, l}
    =
    e^{\pm 2 i \beta}
    \left(
        \tilde{\Delta}_{E, l} \pm i \tilde{\Delta}_{B, l}
    \right)
    \, ,
\end{align}
where the tildes denote quantities without cosmic birefringence.
Thus, cosmic birefringence transforms the polarization power spectra as~\cite{Liu:2006uh,Feng:2004mq}
\begin{align}
    C_l^{E E}
    &=
    \cos^2 (2 \beta) \tilde{C}_l^{E E}
    + \sin^2 (2 \beta) \tilde{C}_l^{B B}
    \,,
    \\
    C_l^{B B}
    &=
    \cos^2 (2 \beta) \tilde{C}_l^{B B}
    + \sin^2 (2 \beta) \tilde{C}_l^{E E}
    \,,
    \\
    C_l^{E B}
    &=
    \frac{1}{2} \sin (4 \beta)
    \left( \tilde{C}_l^{E E} - \tilde{C}_l^{B B} \right)
    \, ,
    \label{eq: ClEB after birefringence}
\end{align}
which can be combined to give $C_l^{EB}=\frac12\tan(4\beta)(C_l^{EE}-C_l^{BB})$~\cite{Zhao:2015mqa}. When the primordial $B$ modes are negligible compared to the $E$ modes, we obtain $C_l^{E B} \simeq \tan(2 \beta) C_l^{E E}$.

In general, $\beta$ depends on $\eta$ through the evolution of $\phi$.
Roughly speaking, there are two main contributions to the CMB polarization power spectra.
One is the reionization epoch contributing to low multipoles, $l \lesssim 10$~\cite{Zaldarriaga:1996ke}, and the other is the recombination epoch contributing to higher $l$~\cite{Bond:1987ub}.
Thus, the $E B$ power spectrum at low and high $l$ is largely determined by the evolution of $\phi$ after the reionization and recombination epochs, respectively~\cite{Liu:2006uh,WMAP:2008lyn,Sherwin:2021vgb,Nakatsuka:2022epj}.

If $\phi$ starts to oscillate after the reionization epoch, $\beta$ is approximately constant and we obtain $C_l^{E B} \propto C_l^{E E}$.
If $\phi$ starts to oscillate before the reionization epoch, the reionization bump in the $E B$ power spectrum at low $l$ is suppressed~\cite{Sherwin:2021vgb,Nakatsuka:2022epj}.
If $\phi$ oscillates during the recombination epoch, the \textit{shape} of the $E B$ power spectrum can be different from that of the $E E$ power spectrum at high $l$~\cite{Finelli:2008jv,Nakatsuka:2022epj}, which is relevant to our study.
The information in the shape of the $EB$ power spectrum thus allows for ``tomographic approach,'' with which we can constrain the mechanism inducing cosmic birefringence or axion parameters~\cite{Sherwin:2021vgb,Nakatsuka:2022epj}.

\section{Early dark energy}
\label{sec: EDE}

EDE is an additional energy component that behaves like dark energy at early times and contributes to the energy density budget around the recombination epoch~\cite{Karwal:2016vyq}.
This contribution increases the Hubble expansion rate around the recombination epoch, which, in turn, reduces the size of the sound horizon~\cite{Bernal:2016gxb}.
As a result, EDE increases the inferred value of $H_0$ from CMB observations and alleviates the Hubble tension.

The EDE energy density should decay faster than the matter density contribution after recombination, so as not to affect the late-time cosmic expansion.
The EDE models that satisfy this property are based on a pseudoscalar field potential given by~\cite{Poulin:2018cxd}
\begin{align}
    V(\phi)
    =
    m^2 f^2 \left[
        1 - \cos \left( \frac{\phi}{f} \right)
    \right]^n
    \, ,
    \label{eq:potential}
\end{align}
where $f$ is the breaking scale of the symmetry related to $\phi$ and $m$ determines the potential scale.
Around $\phi = 0$, this potential is approximated by $V(\phi) \propto \phi^{2n}$.
Thus, once $\phi$ starts to oscillate, its energy density decreases as $\rho_\phi \propto a^{-6n/(n+1)}$~\cite{Turner:1983he}, which means that $\rho_\phi$ decreases faster than the matter density for $n > 1$.
In the following, we choose $n = 3$ as in the previous work~\cite{Herold:2021ksg}.

EDE is usually characterized by three parameters, $(f_\mathrm{EDE}, z_c, \theta_i)$, where $f_\mathrm{EDE}$ is the maximum energy density fraction of EDE at the critical redshift $z_c$, and $\theta_i$ is the initial value of $\theta \equiv \phi/f$.
The previous work~\cite{Herold:2021ksg} has performed a profile likelihood analysis to avoid volume effects in the MCMC analysis and derived a robust constraint of $f_\mathrm{EDE} = 0.072\pm 0.036$ at 68\% C.L.
They have also obtained the best-fitting values of the 6 parameters of the base-$\Lambda$CDM model and $(z_c, \theta_i)$ for fixed values of $f_\mathrm{EDE}$ from the {\sl Planck} CMB and the BOSS full-shape galaxy clustering data. In this paper we adopt their best-fitting values for $f_\mathrm{EDE} = 0.01$, $0.07$, and $0.14$, which are summarized in Table~\ref{tab: best-fit parameters}.
\begin{table}[htbp]
    \centering
    \caption{Best-fitting values of the base-$\Lambda$CDM and EDE model parameters for $f_\mathrm{EDE} = 0.01, 0.07$, and $0.14$.
    }
    \label{tab: best-fit parameters}
    \begin{tabular}{ l | c  c  c }
        \hline
        $f_\mathrm{EDE}$ & 0.01 & 0.07 & 0.14 
        \\
        \hline
        100 $\omega_b$ & 2.248 & 2.259 & 2.276
        \\
        $\omega_\mathrm{cdm}$ & 0.1200 & 0.1260 & 0.1341
        \\
        $100 \, \theta_s$ & 1.042 & 1.042 & 1.041
        \\
        $\ln( 10^{10} A_s )$ & 3.046 & 3.056 & 3.070
        \\
        $n_s$ & 0.9704 & 0.9794 & 0.9927 
        \\
        $\tau_\mathrm{reio}$ & 0.05492 & 0.05492 & 0.05534
        \\
        $\log_{10} z_c$ & 3.551 & 3.548 & 3.560
        \\
        $|\theta_i|$ & 2.752 & 2.758 & 2.769 
        \\
        \hline
    \end{tabular}
\end{table}

\begin{figure}[tbp]
    \centering
    \includegraphics[clip,width=\linewidth]{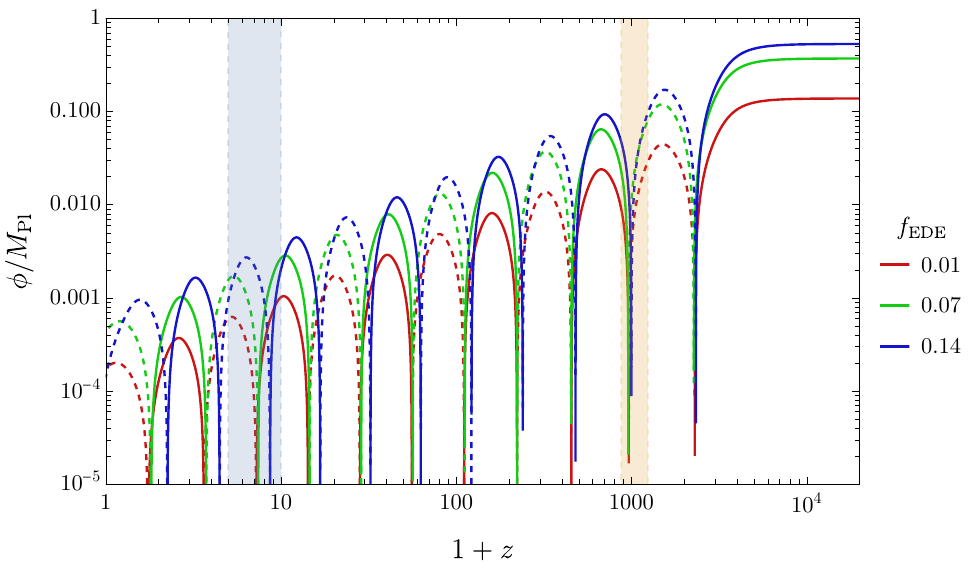}
    \caption{
        Time evolution of $\phi$ for the best-fitting parameters with $f_\mathrm{EDE} = 0.01, 0.07$, and $0.14$ given in Table~\ref{tab: best-fit parameters}.
        The dashed lines represent $-\phi$.
        The orange and blue regions represent the recombination and reionization epochs, respectively.
    }
    \label{fig: EDE evolution}
\end{figure}

We show the time evolution of $\phi$ in Fig.~\ref{fig: EDE evolution}.
The orange and blue regions represent the recombination and reionization epochs, respectively.
If the EDE has a Chern-Simons coupling to photons, the time evolution of $\phi$ induces isotropic cosmic birefringence.
The difference in the EDE parameters affects the amplitude and phase of oscillations.
While the amplitude is proportional to $\beta$ (that is, the amplitude of $C_l^{E B}$), the phase affects the shape of $C_l^{E B}$.

\section{Cosmic birefringence from EDE}
\label{sec: CB from EDE}

The authors of Ref.~\cite{Nakatsuka:2022epj} included the dynamics of $\phi$ in the Boltzmann equation, assuming that the energy density of $\phi$ is negligible in the Friedmann equation and ignoring the gravitational lensing effect of the CMB. In this paper we account for both effects. 

Specifically, we modify the publicly-available \texttt{CLASS\_EDE} code~\cite{Hill:2020osr}, which is based on the linear Boltzmann solver \texttt{CLASS}~\cite{Lesgourgues:2011re,Blas:2011rf} and solves the equation of motion for $\phi$ with the potential given in Eq.~\eqref{eq:potential}, to include $\beta(\eta)$ in the line-of-sight integral solution [Eq.~\eqref{eq: Solution for Boltzmann}] and 
output the $EB$ power spectrum using Eq.~\eqref{eq:cl}.
We include the lensing effect as explained in Ref.~\cite{Naokawa:2022xxx}.

\begin{figure}[tbp]
    \centering
    \includegraphics[clip,width=\linewidth]{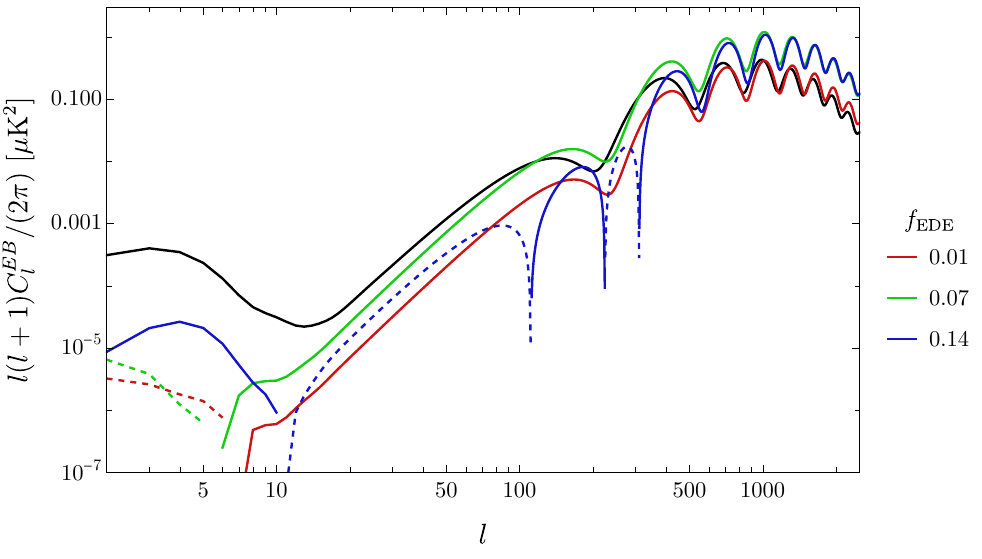}
    \caption{
        $EB$ power spectrum in the EDE model for the best-fitting parameters given in Table~\ref{tab: best-fit parameters} and $g = M_\mathrm{Pl}^{-1}$.
        The colored lines represent the $EB$ power spectra for $f_\mathrm{EDE} = 0.01$, $0.07$, and $0.14$, whereas
        the black line represents the $E E$ power spectrum multiplied by a constant factor for comparison. The solid and dashed lines show positive and negative values, respectively.
    }
    \label{fig: EB main result}
\end{figure}

In Fig.~\ref{fig: EB main result} we show the $E B$ power spectrum induced by the EDE field shown in Fig.~\ref{fig: EDE evolution}.
Here, we fix $g = M_\mathrm{Pl}^{-1}$ by the reduced Planck constant $M_\mathrm{Pl}\equiv (8\pi G)^{-1/2}$. One can rescale our results for arbitrary values of $g$ by $C_l^{E B}=gM_\mathrm{Pl}C_l^{EB}(g=M_\mathrm{Pl}^{-1})$. 
As the sign of $C_l^{E B}$ depends on those of $g$ and $\theta_i$, 
we assume $g > 0$ and $\theta_i > 0$ so that $C_l^{E B} > 0$ at $l \gtrsim 500$.
For comparison, the black line represents $C_l^{E E}$ multiplied by a constant.

\begin{figure}[tbp]
    \centering
    \includegraphics[clip,width=\linewidth]{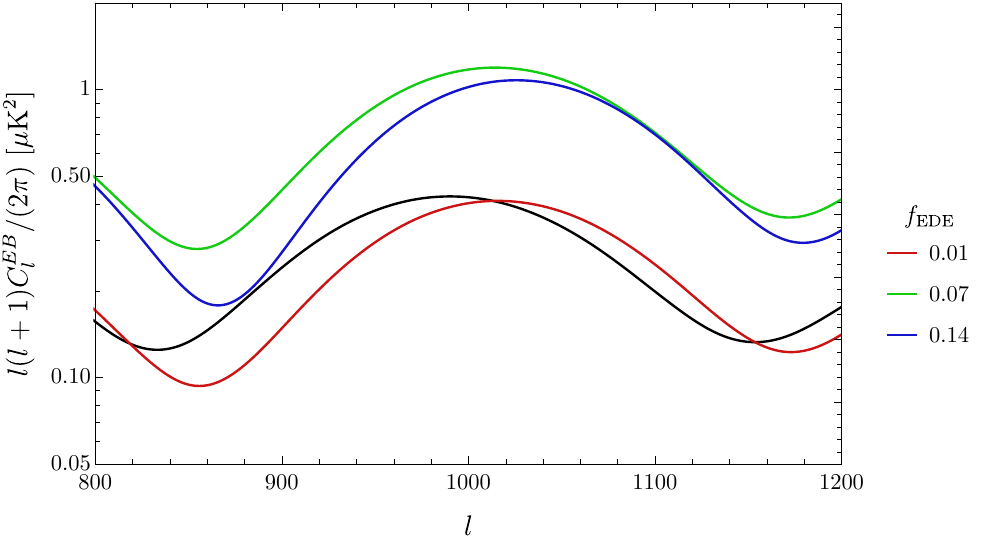}
    \caption{
        $EB$ power spectrum in the EDE model at $800 \leq l \leq 1200$.
        The lines are the same as in Fig.~\ref{fig: EB main result}.
    }
    \label{fig: EB high l}
\end{figure}

We find that $C_l^{E B}$ has a wide variety of shapes, which are different from $C_l^{E E}$ commonly assumed in the data analysis~\cite{Minami:2020odp,Diego-Palazuelos:2022dsq,Eskilt:2022cff,Eskilt:2022wav}.
One can understand the shape of $C_l^{EB}$ from the time evolution of $\phi$.
Since the CMB photons contribute to the power spectrum in a different way depending on the time of last scattering, the time-dependent $\beta$ leads to a modulation of the shape of $C_l^{E B}$ with respect to that of $C_l^{E E}$~\cite{Nakatsuka:2022epj}.
In our case, $\phi$ flips the sign during the recombination epoch as seen in Fig.~\ref{fig: EDE evolution}, and thus $C_l^{E B}$ can also flip the sign at $10 \lesssim l \lesssim 500$.
This effect also shifts the peak positions of $C_l^{E B}$ at high $l$ as seen in Fig.~\ref{fig: EB high l}.
Since the positive contribution to $C_l^{E B}$ comes from the earlier stage of the recombination epoch, the peak shifts to higher $l$.
We can also understand the diverse behaviors of $C_l^{E B}$ at $l \lesssim 10$ from the variation of the oscillation phases during the reionization epoch.

Let us relate our $C_l^{E B}$ to the observed value of $\beta$.
The observational determination of $\beta$ typically assumes a constant rotation angle multiplying $C_l^{EE}-C_l^{BB}$~\cite{Minami:2020odp,Diego-Palazuelos:2022dsq,Eskilt:2022cff,Eskilt:2022wav}. Although our $C_l^{E B}$ is not proportional to $C_l^{E E}-C_l^{BB}$, one may find an approximate proportionality factor at $l \gtrsim 500$ and translate our $C_l^{E B}$ into $\beta$ by comparing the maximum values of 
$C_l^{E E} - C_l^{B B}$ and $C_l^{E B}$.
Considering Eq.~\eqref{eq: ClEB after birefringence}, we define the effective value of $\beta$, $\beta_\mathrm{eff}$, by
\begin{align}
    \mathrm{max} \left[ C_l^{E B} \right]
    \equiv
    \frac{1}{2} \tan ( 4\beta_{\mathrm{eff}}) \, 
    \mathrm{max} \left[ C_l^{E E} - C_l^{B B} \right].
\end{align}
By requiring $\beta_\mathrm{eff} = 0.34^{\circ}$, we estimate $g$ as
\begin{align}
    \frac{g}{M_\mathrm{Pl}^{-1}}
    =
    \left(
        1.2,  
        0.42, 
        0.47  
    \right)
    \, ,
\end{align}
for $f_\mathrm{EDE} = 0.01$, $0.07$, and $0.14$, respectively.
Since the decay constant, $f$, is typically of $\mathcal{O}(0.1 M_\mathrm{Pl})$ for the best-fitting parameters, this estimation implies $g = \mathcal{O}(0.01 \,\text{--}\, 0.1) f^{-1}$.

\begin{figure*}[tbp]
    \centering
    \includegraphics[clip,width=85mm]{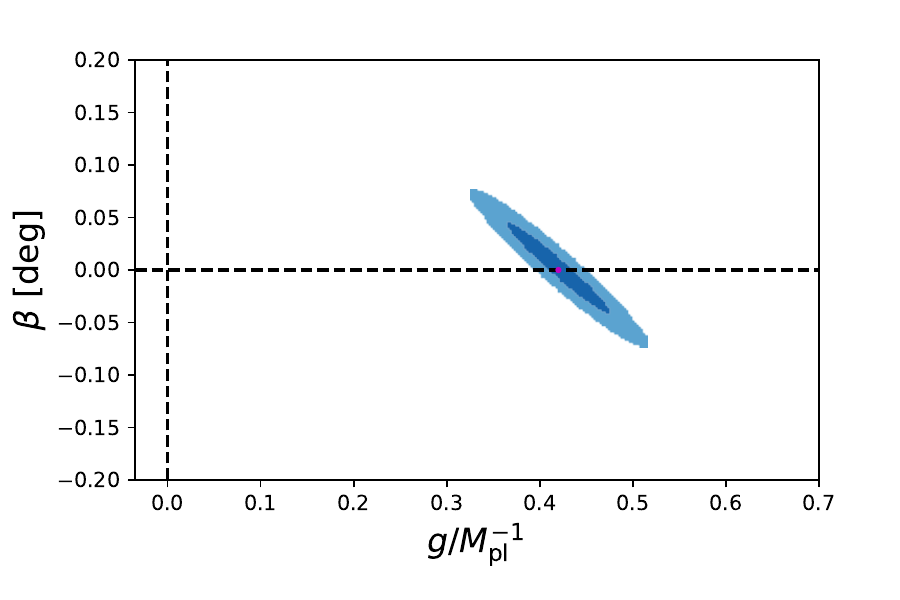}
    \includegraphics[clip,width=85mm]{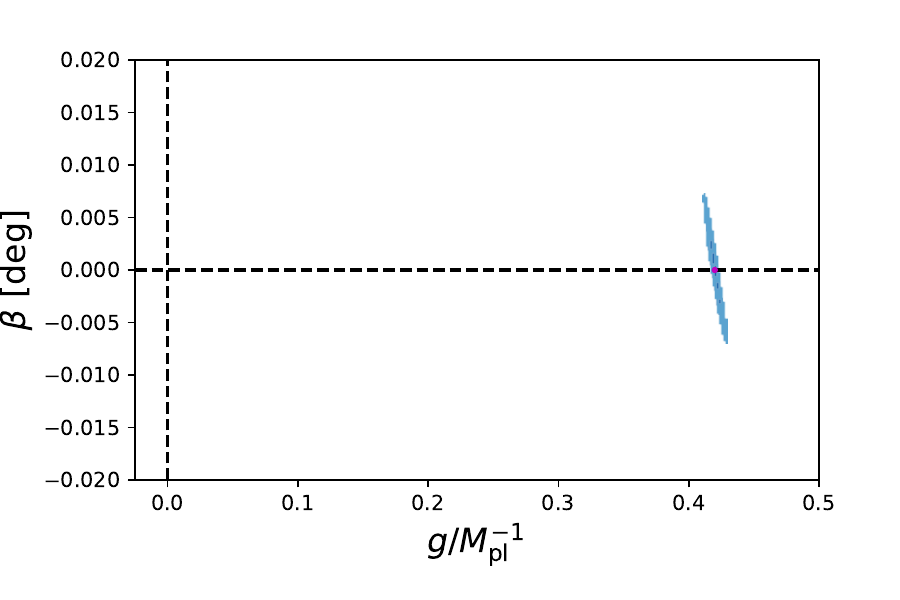}
    \caption{
    The expected 1 and 2\,$\sigma$ error contours on $g/M^{-1}_{\rm Pl}$ and $\beta$ (in units of degrees) for $f_{\rm EDE}=0.07$. The fiducial value of $g/M^{-1}_{\rm Pl}$ is chosen to give $\beta_{\rm eff}=0.34^\circ$ (magenta dots). We assume the SO-like (Left) and S4-like (Right) experiments.
    }.
    \label{fig:const}
\end{figure*}

Finally, we forecast the testability of the EDE model
by measuring the $E B$ power spectrum in future CMB experiments.
The overarching question is whether one can distinguish between the $EB$ power spectra from the EDE model and the simplest form of birefringence by a constant rotation angle $\beta$. To this end, we first compute the following chi-squared \cite{Nakatsuka:2022epj}: 
\begin{align}
    \chi^2(\bm{p}) = f_{\rm sky}\sum_{l=l_{\rm min}}^{l_{\rm max}} (2l+1)\frac{[\hat{C}^{EB}_l-C^{EB,\rm th}_l(\bm{p})]^2}{\hat{C}^{EE}_l\hat{C}^{BB}_l} 
    \,, 
\end{align}
where $\hat{C}^{XY}_l$ is an observed power spectrum, $f_{\rm sky}$ is a sky fraction used for the analysis, $\bm{p}$ is the parameters to be constrained, and $l_{\rm min}$ and $l_{\rm max}$ are the minimum and maximum multipoles included in the analysis, respectively. Here, $C^{EB,\rm th}_l$ is a theoretical model for the $EB$ power spectrum given by
\begin{align}
    C^{EB,\rm th}_l = \frac{g}{M_{\rm Pl}^{-1}}C_l^{EB,0} + \frac{\sin(4\beta)}{2}(\tilde{C}_l^{EE}-\tilde{C}_l^{BB}) 
    \,, 
\end{align}
where $C_l^{EB,0}$ is the $EB$ power spectrum for the EDE model with $f_{\rm EDE}=0.07$ and $g=M_{\rm Pl}^{-1}$. 
For a given observed $\hat{C}^{EB}_l$ we compute
$\chi^2$ for each parameter set, $\bm{p}=(g/M_{\rm Pl}^{-1},\beta)$, and obtain the posterior distribution, $P(\bm{p}|\hat{C}^{EB})\propto\exp\left[-\chi^2(\bm{p})/2\right]$ \cite{Nakatsuka:2022epj}. 
We assume that the mock data, $\hat{C}^{EB}_l$, is described by the EDE model with $f_{\rm EDE}=0.07$ and $g/M_{\rm Pl}^{-1}=0.42$.

Here, we consider two experiments: the Simons Observatory (SO)~\cite{SimonsObservatory:2018koc} and CMB-S4~\cite{CMBS4}.
We choose the same experimental setup for CMB-S4 as described in Ref.~\cite{Nakatsuka:2022epj}, while for SO we use the noise curves provided by the SO collaboration.%
\footnote{\url{https://github.com/simonsobs/so_noise_models}}
In Fig.~\ref{fig:const} we show the expected error contours on $g/M_{\rm Pl}^{-1}$ and $\beta$.
For both SO and CMB-S4, the constant rotation alone ($g/M_{\rm Pl}^{-1}=0$) cannot explain the mock data, and the EDE model will be distinguished from a constant rotation.

Discovery of the feature in the $EB$ spectrum predicted by the EDE model would be a breakthrough in cosmology and fundamental physics.

\section{Summary}
\label{sec: summary}

Polarization of the CMB is a powerful probe of new physics beyond the standard model of elementary particles and fields~\cite{Komatsu:2022nvu}. In this paper we connected two hints of new physics from the current cosmological datasets: isotropic cosmic birefringence and EDE. We included the dynamics of the EDE field, $\phi$, and its coupling to photons in the Boltzmann equation for the CMB polarization and evaluated the $E B$ power spectrum, $C_l^{E B}$, induced by isotropic cosmic birefringence.
The shape of $C_l^{E B}$ (Fig.~\ref{fig: EB main result}) is different from that of $C_l^{EE}$ and depends sensitively on the EDE parameters given in Table~\ref{tab: best-fit parameters}. We understood this result in terms of the time evolution of $\phi$ shown in Fig.~\ref{fig: EDE evolution}.

We roughly translated the obtained $C_l^{E B}$ into an effective rotation angle, $\beta_\mathrm{eff}$.
By requiring that $\beta_\mathrm{eff}$ is equal to the observed rotation angle, we found $g = \mathcal{O}(0.01 \,\text{--}\, 0.1) f^{-1}$.
We also discussed the testability of the EDE model and showed that the cosmic birefringence predicted by the EDE model can be distinguished from that with a constant rotation angle by measuring $C_l^{E B}$ in future CMB experiments such as SO and CMB-S4.

Our analysis can also be applied to other EDE models~\cite{Agrawal:2019lmo,Ye:2020btb}.
Since the shape of $C_l^{E B}$ is sensitive to the oscillation phase of the EDE field during the recombination and reionization epoch, the shape of $C_l^{E B}$ can be different in other models.
Although we have focused on the effect of isotropic cosmic birefringence on $C_l^{E B}$, the perturbations of the EDE field will induce the anisotropic cosmic birefringence~\cite{Capparelli:2019rtn}, which may provide a distinct signature in $C_l^{E B}$.
We leave this to future work.

\begin{acknowledgments}
We thank E.~G.~M. Ferreira and L. Herold for their help with the EDE parameters and \texttt{CLASS\_EDE} code, and the participants of the workshop YITP-T-21-08 on ``Upcoming CMB observations and Cosmology'' for useful discussion. This work was supported in part by the Program of Excellence in Photon Science (K.M.),
the Forefront Physics and Mathematics Program to Drive Transformation (FoPM), a World-leading Innovative Graduate Study (WINGS) Program, the University of Tokyo (F.N.), and JSPS KAKENHI Grant No. JP20J20248 (K.M.), No. JP20H05850 (E.K.), No. JP20H05859 (E.K. and T.N.), and No. JP22K03682 (T.N.). Part of this work uses resources of the National Energy Research Scientific Computing Center (NERSC). The Kavli IPMU is supported by World Premier International Research Center Initiative (WPI Initiative), MEXT, Japan.
\end{acknowledgments}

\bibliographystyle{apsrev4-1}
\bibliography{Ref}

\end{document}